# Vortical Reflection and Spiraling Fermi Arcs with Weyl Metamaterials

Hua Cheng[1,2,*], Wenlong Gao[1,*,†], Yangang Bi[1,3,*], Wenwei Liu[2], Zhancheng Li[2], Qinghua Guo[1], Yang Yang[1], Oubo You[1], Jing Feng[3], Hongbo Sun[3,4], Jianguo Tian[2], Shuqi Chen[2,†], Shuang Zhang[1,†]

[1]School of Physics & Astronomy, University of Birmingham, Birmingham B15 2TT, UK.

[2]The Key Laboratory of Weak Light Nonlinear Photonics, Ministry of Education, School of Physics, TEDA Institute of Applied Physics, and Renewable Energy Conversion and Storage Center, Nankai University, Tianjin 300071, China.

[3]State Key Lab of Integrated Optoelectronics, College of Electronic Science and Engineering, Jilin University, Changchun 130012, China

[4]State Key Laboratory of Precision Measurement Technology & Instruments, Department of Precision Instrument, Tsinghua University, Beijing 100084, China

*These authors contributed equally to this work.

†Corresponding author. Email: bhamwxg334@gmail.com (W.G); schen@nankai.edu.cn (S.C.); s.zhang@bham.ac.uk (S.Z.)

**Scatterings and transport in Weyl semimetals have caught growing attention in condensed matter physics, with observables including chiral zero modes and the associated magnetoresistance and chiral magnetic effects. Measurement of electrical conductance is usually performed in these studies, which, however, cannot resolve the momentum of electrons, preventing direct observation of the phase singularities in scattering matrix associated with Weyl point. Here we experimentally demonstrate a helical phase distribution in the angle (momentum) resolved scattering matrix of electromagnetic waves in a photonic Weyl metamaterial. It further leads to spiraling Fermi arcs in an air gap sandwiched between a Weyl metamaterial and a metal plate. Benefiting from the alignment-free feature of angular vortical reflection, our findings establish a new platform in manipulating optical angular momenta with photonic Weyl systems.**

Weyl points are linear band crossings at discrete points in the momentum space, corresponding to three-dimensional (3D) extensions of the two-dimensional (2D) Dirac nodes, and functioning as monopoles of Berry flux with topological charges defined by the Chern numbers. In photonics, Weyl points have been realized in systems with broken inversion symmetry such as photonic crystals and metamaterials [1–10], and with broken time reversal symmetry such as magnetized plasma [11,12] and in synthetic dimensions [13,14]. As a key signature of Weyl systems, Fermi arcs that connect between the projections of Weyl points of opposite topological charges have been observed [15–17]. In phononics, the topological features of the Fermi arcs have been utilized to realize negative refractions [18]. However, the manifestation of the topological nature of Weyl points in their interaction with free space radiations remain obscure in both quantum and classical systems.

Scatterings and transport in Weyl semimetals have caught growing attention in both condensed matter and classical physics. In condensed matter systems, the chiral zero modes and the associated magnetoresistance and chiral magnetic effects underlie some highly unusual scattering and transport properties of Weyl semimetals [19–26]. It has also been proposed that Weyl semimetals may have handedness dependent Imbert–Fedorov shift within the bulk Weyl material [27]. Electrical conductance is usually measured in the experiments, which has contribution from both the bulk Weyl nodes and the fermi arcs [25]. However, it is generally difficult to perform momentum resolved scattering and transport studies, since the leads that transport the electrons to the Weyl semimetals, by their nature, cannot control the momentum of the input electrons. Therefore, previous attempts to associate the topology to scattering matrix mostly remain theoretical [28,29].

In this work, vortical phase profile in the momentum space is demonstrated for spin polarized electromagnetic waves reflected by a photonic Weyl system, which represents a key signature of the topological nature of the Weyl points and could lead to alignment-free phase plates for vortex beams generation. In contrast, conventional vortex beam generation relies on spatially varying phase elements, such as liquid-crystal q-plates, spatial light modulators, and metasurfaces [30,31], including the Pancharatnam–Berry phase metasurfaces that utilize subwavelength anisotropic phase elements to impose additional phase on spin-flipped circular polarization states [32–35]. All the above-mentioned devices contain inhomogeneous patterns in the real space, requiring stringent alignment of the incident optical beam to the defect center of the phase profile, though an interlaced metasurface has been proposed to relax the stringent alignment to some extent [36]. We further discover a novel type of spiraling guided mode in a hybrid waveguide system formed by the Weyl metamaterial and a metal plate, enabled by the vortical phase profile in the momentum space. These observations point to new applications of photonic Weyl systems for manipulation of the angular momenta of electromagnetic waves in both free space and guided systems.

A general sketch of the alignment free vortex beam generation by Weyl metamaterial is given in Fig. 1a. An input beam with spin angular momentum is incident onto a periodic metamaterial at normal incidence. The reflected beam is converted to the opposite spin angular momentum and carries an extra orbital angular momentum. Benefiting from the translational invariance nature of the Weyl metamaterial, the incident beam is not required to be aligned to a certain position on the metamaterial. We consider a photonic Weyl system consisting of four Weyl nodes – a minimum number for systems respecting time reversal symmetry. The Weyl metamaterial is constructed by stacking of metallic saddle shaped connective coil into a 3D array. The metallic inclusions are arranged in the *u*, *v*, and *w* directions with periodicity of $p_u = p_w = 3$ mm and $p_v = 4.5$ mm. The *uv*-plane forms an angle of 45° with the interface of the metamaterial, which lie in the horizontal plane (*yz*-plane). The metallic inclusions are embedded in a dielectric medium with dielectric constant of about 2.2. The unit cell of the metallic saddle shaped connective coil possesses $D_{2d}$ node group symmetry (Fig. 1b), in which four ideal Weyl nodes exist on the transverse $C_2$ high symmetry line [10], as shown in Fig. 1c. The 2D band structures of $S_1$ and $S_2$ planes (parallel to *yz*-plane as shown in Fig. 1c) are calculated and overlapped in the same plot (Fig. 1d). In the *yz*-plane, four Weyl nodes project into three points onto the surface Brillouin zone, with the one at the center possessing a topological charge of 2.

In Fig. 1e, where the positive/negative Weyl nodes are shown as the red/blue spheres with arrows indicating the directions of the Berry curvatures, we define a cylindrical surface that contains only the positive Weyl points while excluding the negative ones. The integral of Berry curvatures across the surface of the cylinder, defined as the Chern number, is given by the total number of Weyl points it encloses, which is 2. Imagine we unwrap the cylinder to form an effective planar 2D Brillouin zone shown in Fig. 1f, due to the non-zero Chern number, this is effectively a 2D Chern insulator with insulating bulk and conducting channels. The nontrivial topology of Chern insulator can not only be revealed by the number of surface states, but also by the scattering matrix S [37,38], whose topological index is formulated as: $W = \frac{1}{2\pi i}\int_0^{2\pi} dk \frac{d}{dk} \log \det(S(k))$, i.e. the total winding number in $2\pi$ of the phase of the eigenvalues of the scattering matrix. It has been rigorously proven that the topological classification of the Hamiltonian and the scattering matrix are equivalent [29]. The scattering matrix method however, are more accessible for investigation in photonics with the readily available components to perform the angle resolved measurements.

We employ a microwave reflection setup for the measurement of angle-resolved reflection phase, as illustrated in Fig. 2a, wherein two microwave horn antennas are used as a transmitter and a receiver respectively. The S-parameters of the four combinations of s and p polarizations are measured for varying azimuthal angle θ, corresponding to a circle of radius $k_0\sin(\varphi)$ in the surface Brillouin zone, where $\varphi$ is the incident elevation angle. Generally, for the quantum states located on a cylinder that has an arbitrary radius enclosing the Weyl nodes, the classification is the A class which has no symmetries in the Wigner-Dyson topological classification [37]. In such case, the scattering matrix can be classified by integers Z. In our photonic system the scattering is characterized by Jones matrix, a 2-by-2 matrix with TE and TM as the two scattering channels. The two Weyl nodes projected onto the center of surface Brillouin give rise to a nontrivial Jones matrix with Z=2, corresponding to sum of their topological charges, expressed as $Z = \frac{1}{2\pi}\sum_{i=1,2}\int d\phi_i(\theta)$ where $\phi_i$ are the phases of Jones matrix's eigen values. By measuring the Jones matrix for different incident azimuth angles surrounding the surface normal and retrieving the reflection phases for the two eigen polarization states, we experimentally confirm the intriguing phase windings in reflection around the projected Weyl point, as shown by Fig. 2b. The measured reflection phase agrees well with the analytical and numerical results even though the eigen-polarizations are slightly affected by the dissipation according to our simulations (Supplementary information 1).

To understand the mechanism of the windings of the reflection phases in the k-space within classical electromagnetic wave theory, we apply the effective medium theory to describe the Weyl metamaterial. we can derive the scattering matrix and the corresponding reflection phases as (Supplemental information 2):

$$\phi_1(\theta) = \phi_2(\theta+\pi) = -i\ln \frac{(\frac{\eta_z}{\eta_d} - \frac{\eta_y \eta_d}{\eta_z^2})\sin\theta \pm i\sqrt{(1-\frac{\eta_y}{\eta_z})^2 \cos^2\theta + 4\frac{\eta_y}{\eta_z}}}{(\frac{\eta_z}{\eta_d} + \frac{\eta_y \eta_d}{\eta_z^2})\sin\theta - i(1+\frac{\eta_y}{\eta_z})\cos\theta} \quad (1)$$

Where $\eta_{y,z} = \sqrt{\mu_{y,z}/\varepsilon_{y,z}}$ is the impedance of the Weyl metamaterial in y and z direction and $\eta_d =$

$\sqrt{\mu_d/\varepsilon_d}$ the impedance of the incident medium. The corresponding eigen-polarizations are two orthogonal linear polarization states whose orientations change by $\pi$ when the incident azimuth angle goes through a full turn, as shown in Fig. 2b. If the Weyl frequency is far from the resonance, meaning $\eta_z = \eta_y$, equation (1) can be simplified as: $\phi_1(\theta) = \phi_2(\theta + \pi) = -i \ln \frac{\eta_z + i\eta_d \tan\frac{\theta}{2}}{\eta_z - i\eta_d \tan\frac{\theta}{2}}$. In an ideal case where the metamaterial and the air are impedance matched to each other, i.e. $\eta_z = \eta_y = \eta_d = 1$, equation (1) could be further simplified as: $\phi_1 = -\theta$, $\phi_2 = -\theta - \pi$, representing two straight lines, with each one linearly winding $2\pi$ phase across a full turn of the incident azimuth angle. In this ideal case, the scattering matrix $\hat{S}$ around the Weyl point is given by (Supplemental information 2):

$$\hat{S} = e^{-i\theta} \begin{bmatrix} \cos\theta & \sin\theta \\ \sin\theta & -\cos\theta \end{bmatrix} \quad (2)$$

This scattering matrix operated on circularly polarized states: $|R\rangle = [1, -i]^T, |L\rangle = [1, i]^T$ give $\hat{S}|R\rangle = e^{-i2\theta}|L\rangle$ and $\hat{S}|L\rangle = |R\rangle$. Note that the right-handed circular polarization state $|R\rangle$ is converted to the opposite left-handed circular polarization state $|L\rangle$ with an extra $2\theta$ phase, while the conversion from $|L\rangle$ to $|R\rangle$ keeps the phase unchanged. It follows that a Gaussian beam with polarization state $|R\rangle$ would be converted into a vortex beam with orbital angular momentum of 2 upon reflection (supplementary information 3).

For the realistic Weyl metamaterial designed here, despite the presence of impedance mismatch with air that leads to eigen states of non-circular polarizations and a nonlinear dependence of the reflection phases over the azimuth angle (Supplementary information 4), the same topological feature is maintained, i.e an overall winding phase of $2\pi$ for both eigen states [Fig. 2(b)]. Our calculations show that for an incident beam with the right-handed elliptical polarization state described in Fig. 2(c), the Weyl metamaterial would give rise to a winding reflection phase profile given in Fig. 2(d), which fits well with the numerical simulation. The slight differences between the numerical and experimental results can be attributed to the loss in the metamaterial and the deviation of the radiation of horn antennas from a plane wave in the experiment. The details of the experimental setup for the measurement of reflections can be found in Supplementary Information Note 8.

Based on the vortical reflection of the scattering matrix around the Weyl points, we further observe spiraling Fermi arcs in a guided system. In the momentum space, Fermi arcs connect projections of the Weyl nodes with opposite chirality on the surfaces. The detailed "path" of each fermi arc is subject to the specific boundary configurations. It was recently theoretically demonstrated that Fermi arc surface states may give rise to a spiraling pattern in presence of descending potential barrier on the Weyl semimetal's surfaces [39,40]; however, there has been no feasible schemes to realize such potential barrier and consequently the spiraling fermi arcs. Here we demonstrate that the fermi arcs possessing arbitrary number of windings around the projected Weyl points can be realized in a photonic Weyl semimetal-waveguide hybrid system, revealing a new intriguing physical properties of photonic Weyl systems. The metamaterial-waveguide hybrid system is formed by an air layer sandwiched between a metallic ground plane and an ideal Weyl metamaterial. In such a system, the surface arc states exponentially decay into the Weyl metamaterial, while forming waveguide modes in the thin air layer between the Weyl metamaterial and the confining metallic plate. By measuring the electromagnetic field distribution inside the waveguide, we are able to visualize the Fermi arc's iso-frequency contour in the momentum space through Fourier transformation, and demonstrate the windings of the Fermi arcs in the vicinity of the Weyl frequency.

Spiraling of the Fermi arcs is a direct consequence of the interplay between the nontrivial reflection phases $\phi_1, \phi_2$ (Eq. 2) of the Jones matrix (Eq. 3) of the Weyl nodes and the waveguide dispersion. As depicted in Fig. 3(a), the round-trip propagation phase of the air gap together with the reflection phase of the PEC boundary can be expressed as $2k_x d + \pi = 2d\sqrt{k_0^2 - k_r^2} + \pi$ ($d$ the distance between the PEC boundary and the Weyl semimetal), which only depends on the wave vector $k_x$ when $d$ is fixed, as shown in Fig. 3(a). To form waveguide modes, it is required that the two reflection phases (metamaterial and PEC) and the round-trip propagation phase of the waveguide add up to $2\pi n$. For an ideal system with impedance match between the metamaterial and air, the simple reflection phase $\phi = -\theta$ and $\phi = -\theta - \pi$ leads to spiraling functions $k_r = \sqrt{k_0^2 - \frac{(\theta-\pi)^2}{4d^2}}$ and $k_r = \sqrt{k_0^2 - \frac{\theta^2}{4d^2}}$. Schematic illustration of spiral surface states with projected Weyl nodes in the $k_y$-$k_z$ plane are shown in Fig. 3b. In the realistic metamaterial the spiraling functions are slightly modified due to the nonlinear dependence of the reflection phase over the azimuth angle, but the main features are maintained. Note that the topological charge only determines the number of spiraling fermi arcs, while the windings of the Fermi arcs are largely controlled by the waveguide thickness. When $d$ goes to infinity, the waveguide mode would develop into the bulk continuum of the photon states in vacuum. Polarization in the waveguide of the Fermi arcs are eigen-states of the scattering matrix of the Weyl metamaterial and shows polarization angle rotations along the spiraling Fermi arcs. (Supplementary Information 5)

To experimentally observe the spiraling Fermi arcs, a microwave near-field scan system is employed to detect the spiral Fermi arcs inside the waveguide, as shown by Fig. 3(c) (For details, see Supplementary Information 6 and 7). A series of equi-frequency contours between 12.6 GHz and 13.7 GHz are measured (Fig. 3d) and simulated (Fig. 3e) at a waveguide height of 7 mm. There exist two Fermi arcs that connect to the center projection of Weyl points as expected. At all the frequencies, the measurement clearly shows counterclockwise winding of the Fermi arcs around the center Weyl projection, which match reasonably well with the simulation results in Fig. 3d. With the increase of the frequency, there is a general trend that the Fermi arcs spin around the projected Weyl cone, which is consistent with our previous observation of helicoid surface state around photonic Weyl points [41]. In order to investigate the dependence of the windings of the spiral Fermi arcs upon the waveguide heights, further measurement and simulation are carried out for an increased waveguide height of 11.5 mm (Fig. 3f and g). It is clearly observed that the increased waveguide height leads to larger windings of the Fermi arc around the Weyl point, meanwhile the Fermi arcs become more closely spaced.

To summarize, we have demonstrated the winding phase of scattering matrix around the projected Weyl points in a photonic Weyl metamaterial. With this novel effect, the photonic Weyl metamaterial is found to be a high-performance alignment-free vortex beam phase plate for incident beam of given elliptical polarization. We further designed a Weyl metamaterial-waveguide hybrid system, in which the interplay between the intriguing reflection phase profile surrounding the Weyl point and the dispersion of the waveguide mode renders a twisted Fermi arcs into a spiral form. Our work reveals a promising application of the photonic Weyl metamaterials and a rare manifestation of the spiraling structure in the momentum space, which may be applied for unconventional control of wave propagation in photonic systems.

[40]  S. Li and A. V. Andreev, Phys. Rev. B - Condens. Matter Mater. Phys. **92**, 201107 (2015).

[41]  B. Yang, Q. Guo, B. Tremain, R. Liu, L. E. Barr, Q. Yan, W. Gao, H. Liu, Y. Xiang, J. Chen, C. Fang, A. Hibbins, L. Lu, and S. Zhang, Science (80-. ). **359**, 1013 (2018).10

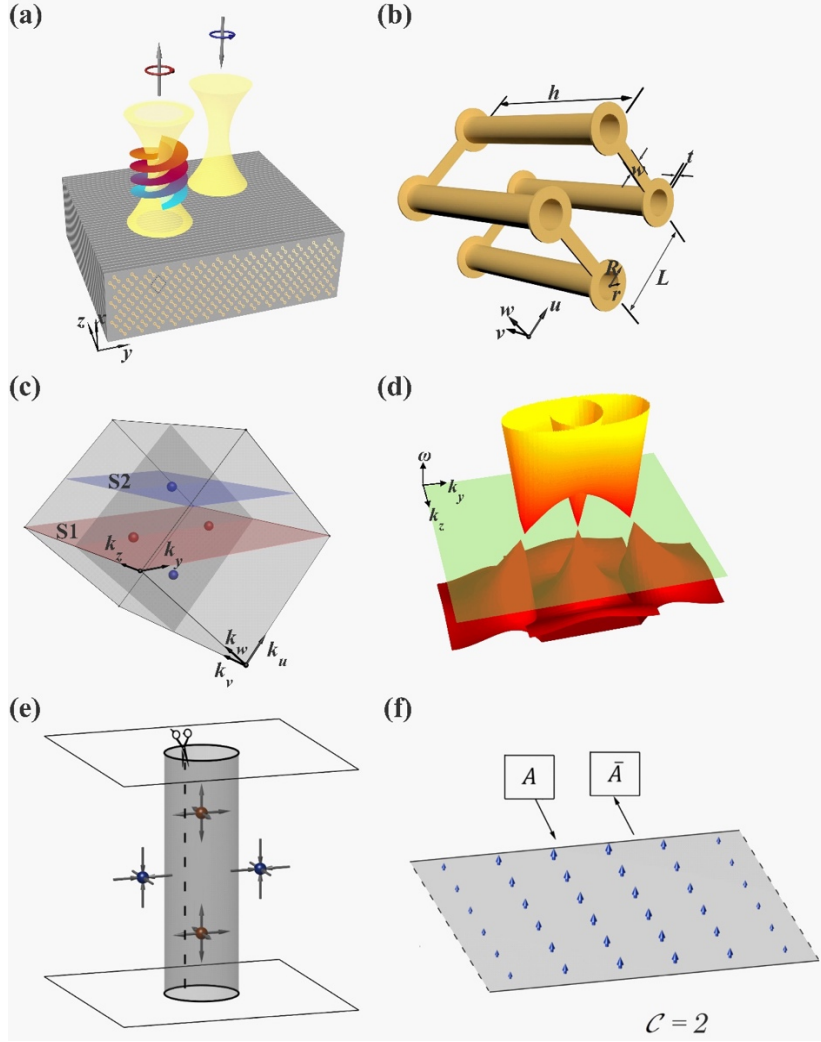

**Fig. 1 | Mechanism of the spiraling behavior of the vortical mirror.** (a) The sketch of the vortical mirror in Weyl metamaterial. The unit cell of the metallic saddle shaped structure with $L = 1.5$ mm, $h = 3$ mm, $R = 0.36$ mm, $r = 0.2$ mm, $w = 0.16$ mm, and $t = 0.035$ mm. Due to the topological scattering phase surrounding the Weyl points, an incident Gaussian beam can be reflected into a vortex beam. (c) Bulk Brillouin zone with four Weyl nodes and their projection on the $k_y$-$k_z$ plane. The top Weyl node (blue) is in the plane of $S_1$ and the middle two Weyl nodes (red) are in the plane of $S_2$. $S_1$ at $k_x=0$ and $S_2$ at $k_x = 0.4\,\pi/a_x$ are both parallel to the $k_y$-$k_z$ plane.(d) 2D dispersion diagram of the $S_1$ and $S_2$ plane. (e) The sketch of a simplest time reversal invariant Weyl metamaterial (f) The effective 2D Brillouin by cutting the cylinder surface in 3D. The Chern number on the 2D Brillouin zone can be represented by the winding number in the scattering matrix linking the input state represented by '$A$' and the output states '$\bar{A}$'

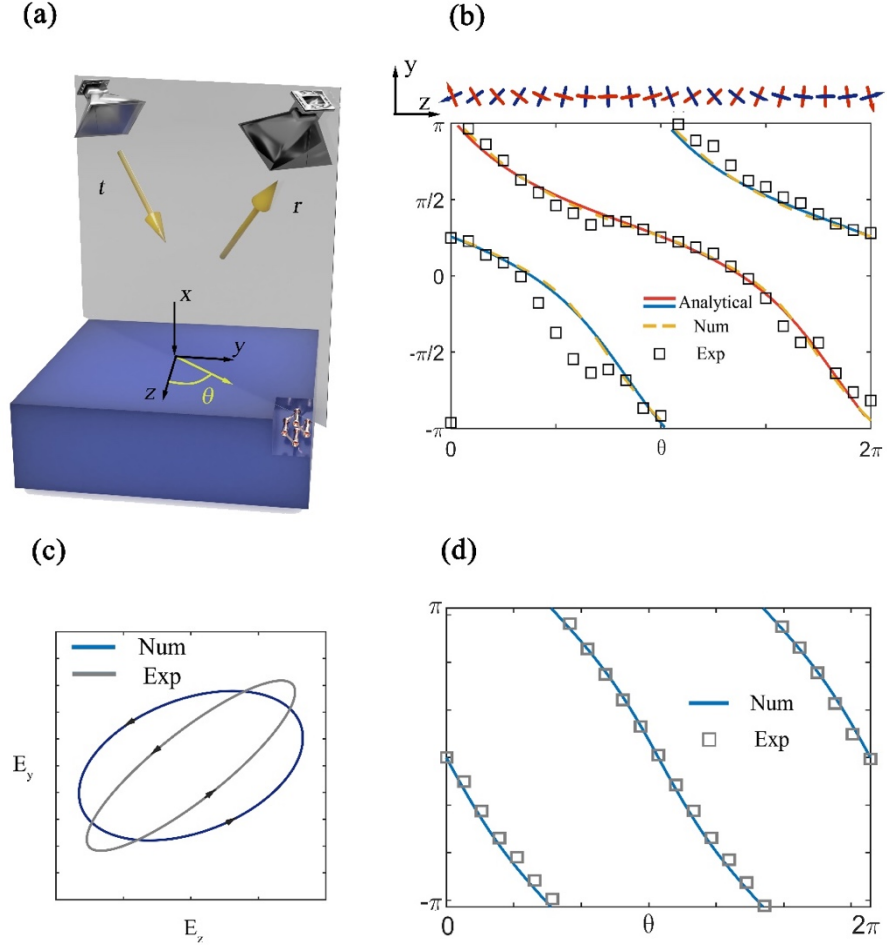

**Fig. 2 | Measurement of the vortical phase mirror in Weyl metamaterial.** (a) The Sketch of the reflection measurement setup in microwave. Letters $t$ and $r$ stand for the transmitter and receiver horn antennas. $\theta$ is the angle between the reflection measurement plane and the z axis. (b) Analytical, numerical and experimental reflection phases of the Weyl metamaterial and the associated eigen-polarizations of the Jones matrix for light incident at different azimuth angles with a fixed elevation angle of 45°. Note that each eigenstate's reflection phase winds $2\pi$, and the eigen-polarizations are linear, and wind $\pi$ for a full turn of rotation. (c)Numerical and Experimental results of the measured $4\pi$ phase winding of the phase mirror. (d) Phase of the reflected vortex beam.

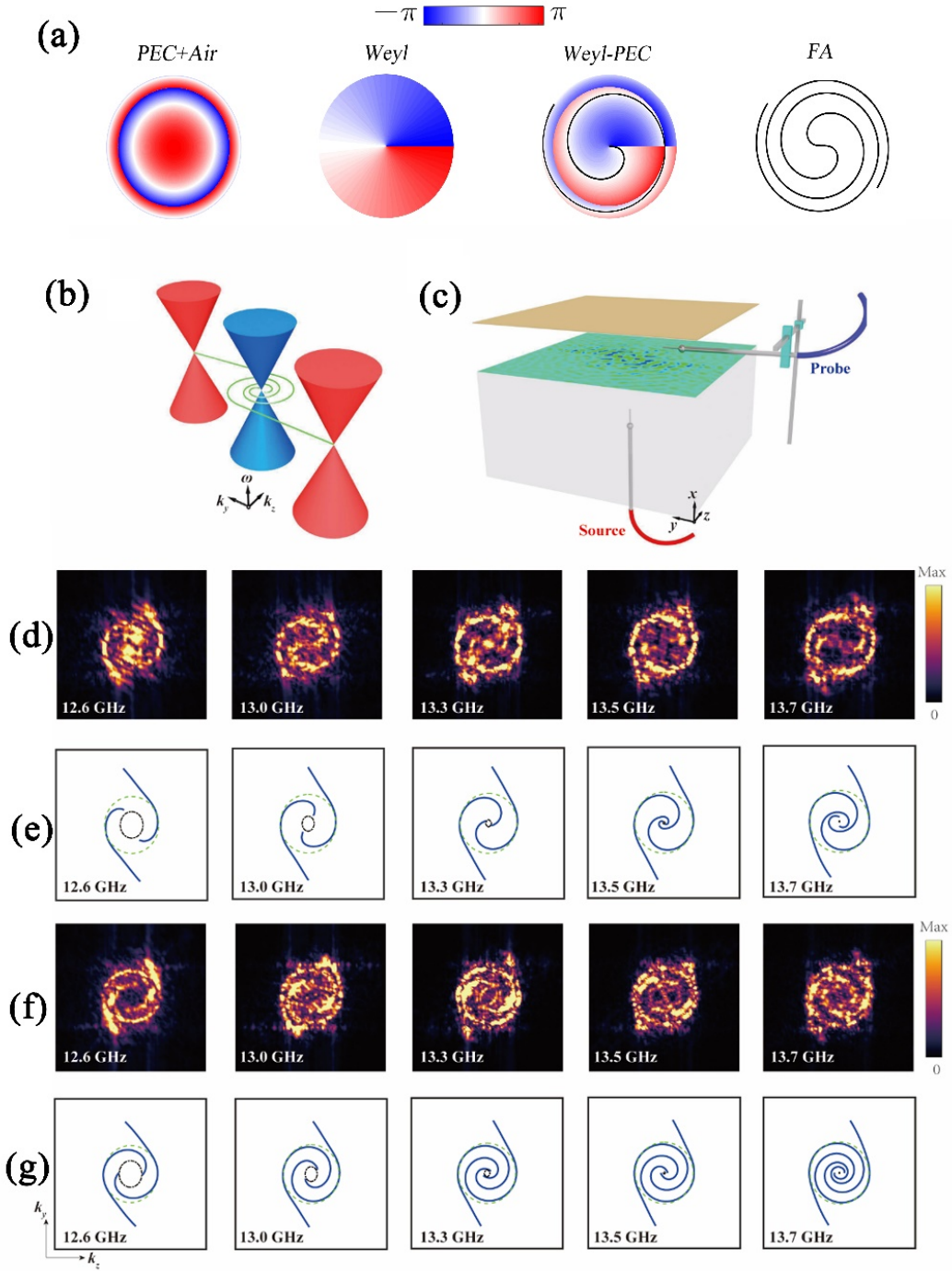

**Fig. 3 | Experimental observation of spiral Fermi arcs.** (a) Illustration of the formation of spiraling Fermi arcs due to the reflection phase by Weyl metamaterial and the PEC. (b) Schematic illustration of spiral surface states with projected Weyl nodes in the $k_z$-$k_y$ plane. (c) Illustration of near-field scanning system, where the source (red) is positioned inside the sample near the center of the top surface, and the probe (blue) scans the top surface inside the waveguide. (d) and (f) Equi-frequency contour measured inside the waveguide from 12.6 GHz to 13.7 GHz, with waveguide height of 7 mm in (d) and 11.5 mm in (f). (e) and (g) Bulk (black dash-dotted) and waveguide modes (blue solid)

simulated by Comsol Multiphysics, with waveguide height of 7 mm in (e) and 11.5 mm in (g). The green dashed circle indicates the air equi-frequency contour. The plotted range for the panels along $k_z$ and $k_y$ are $[-\pi/a_i, \pi/a_i]$, $i = z, y$ respectively.